\documentclass[letter]{aa}
\usepackage{txfonts,epsfig}

\begin{document}
\title{The remnant of Supernova 1987A resolved at 3 mm wavelength}
\author{Ma\v{s}a Laki\'cevi\'c\inst{1,2}
\and    Giovanna Zanardo\inst{3}
\and    Jacco Th.\ van Loon\inst{2}
\and    Lister Staveley-Smith\inst{3,6}
\and    Toby Potter\inst{3}
\and    C.-Y.\ Ng\inst{4}
\and    B.M.\ Gaensler\inst{5,6}
}
\institute{European Organization for Astronomical Research in the Southern
           Hemisphere (ESO), Karl-Schwarzschild-Str.\ 2, D-85748, Garching b.\
           M\"unchen, Germany, \email{mlakicev@eso.org}
\and       Astrophysics Group, Lennard-Jones Laboratories, Keele University,
           Staffordshire ST5 5BG, UK
\and       International Centre for Radio Astronomy Research (ICRAR), M468,
           University of Western Australia, Crawley, WA 6009, Australia
\and       Department of Physics, McGill University, Montr\'eal, QC H3A 2T8,
           Canada
\and       Sydney Institute for Astronomy (SIfA), School of Physics, The
           University of Sydney, NSW 2006, Australia
\and       ARC Centre of Excellence for All-sky Astrophysics (CAASTRO)}
\date{Submitted: 27 January 2012;
      Final version: 19 March 2012}
\abstract
{The proximity of core-collapse Supernova 1987A (SN\,1987A) in the Large
Magellanic Cloud (LMC) and its rapid evolution make it a unique case study of
the development of a young supernova remnant.}
{We aim at resolving the remnant of SN\,1987A for the first time in the 3-mm
band (at 94 GHz).}
{We observed the source at 3-mm wavelength with a 750-m configuration of the
Australia Telescope Compact Array (ATCA). We compare the image with a recent
3-cm image and with archival X-ray images.}
{We present a diffraction-limited image with a resolution of
$0\rlap{.}^{\prime\prime}7$, revealing the ring structure seen at lower
frequencies and at other wavebands. The emission peaks in the eastern part of
the ring. The 3-mm image bears resemblance to early X-ray images (from
1999--2000). We place an upper limit of 1 mJy (2 $\sigma$) on any discrete
source of emission in the centre (inside of the ring). The integrated flux
density at 3 mm has doubled over the six years since the previous observations
at 3 mm.}
{At 3 mm -- i.e.\ within the operational domain of the Atacama Large
Millimeter/submillimeter Array (ALMA) -- SN\,1987A appears to be dominated by
synchrotron radiation from the inner rim of the equatorial ring, characterised
by moderately-weak shocks. There is no clear sign of emission of a different
nature, but the current limits do not rule out such component altogether.}
\authorrunning{Laki\'cevi\'c et al.}
\titlerunning{Remnant of Supernova 1987A resolved at 3 mm}
\maketitle
\section{Introduction}

Supernova 1987A, located in the Large Magellanic Cloud (LMC), was first
observed on 1987 February 24. Since then the evolution of the supernova
remnant (SNR) has been followed across the electro-magnetic spectrum, except
for the far-infrared (FIR) and high radio frequencies. In the radio, since
$\sim$ day 1200 the remnant has been continuously monitored in the 1.4--9 GHz
frequency range with the Australia Telescope Compact Array (ATCA). At these
frequencies, the flux density of the remnant has been steadily increasing over
time (Manchester et al.\ 2002), in an exponential fashion since $\sim$ day
5000 (Zanardo et al.\ 2010). This reflects the propagation of shocks through
the magnetised plasma resulting in synchrotron radiation. However, it is as
yet unclear what is happening at the high-frequency end of the radio spectrum,
where other emission may become apparent including possibly that associated
with dust or an emerging pulsar.

Warm dust ($\approx180$ K) in an equatorial ring
$\approx1\rlap{.}^{\prime\prime}6$ in diameter is linked to mass loss from the
progenitor star (Bouchet et al.\ 2006; Dwek et al.\ 2010). Recently, Matsuura
et al.\ (2011) detected a second dust component, of only $\approx18$ K, which
is also seen at 0.87 mm wavelength (Laki\'cevi\'c et al.\ 2011). Matsuura et
al.\ argued that the cold dust had formed in the ejecta of SN\,1987A. Their
observations did not have sufficient angular resolution to prove this beyond
doubt. While Laki\'cevi\'c et al.\ (2012) constrained the sub-mm emission to
originate from within a few arcseconds from the explosion site, they also
offer scenarios in which the dust might not reside in the ejecta.

The remnant of SN\,1987A (hereafter SNR\,1987A) was first successfully
observed at 3 mm wavelength in 2005 with the ATCA on short baselines. The
remnant was unresolved in the $5^{\prime\prime}$ beam; the integrated flux
density was $11.2\pm2.0$ mJy (Laki\'cevi\'c et al.\ 2011). We have now
attempted to spatially resolve the remnant at this wavelength. Due to the
combination of increased brightness and the ATCA upgrade with the Compact
Array Broadband Backend (CABB; Wilson et al.\ 2011), the SNR has been resolved
for the first time at 3 mm (94 GHz).

The ability to resolve the remnant at high radio frequencies is crucial in
investigating the different sources of emission within the remnant, and as
such the new ATCA observations are a useful complement to future observations
with the Atacama Large Millimeter/submillimeter Array (ALMA).

\section{Observations and data reduction}

%
%
\begin{figure}
\centerline{\psfig{figure=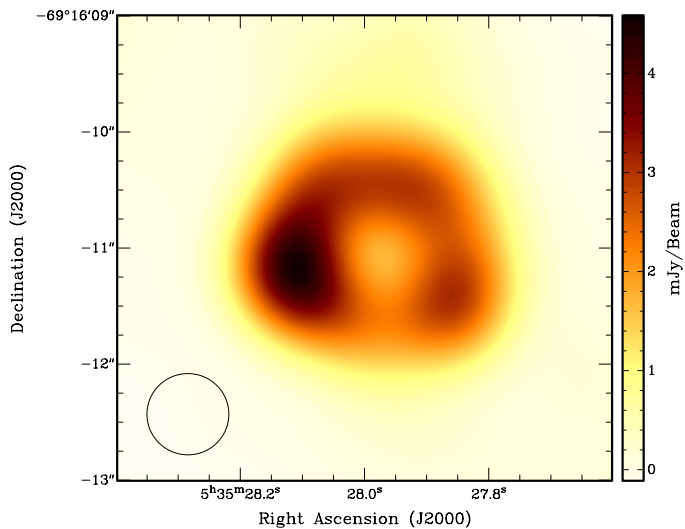,width=90mm}}
\caption[]{Diffraction-limited Stokes-I continuum image of SNR\,1987A at 3 mm
from observations made on 2011 June 30, July 1 and August 2. The image is
restored to a $0\rlap{.}^{\prime\prime}7$ circular beam (plotted in the lower
left corner). The off-source r.m.s.\ noise is $\approx0.5$ mJy beam$^{-1}$.}
\end{figure}

SNR\,1987A was observed at 3-mm wavelength with the ATCA over three 12-hr
sessions, on 2011 June 30, July 1 and August 2 (programme C2495; P.I.\ J.van
Loon) -- days 8892, 8893 and 8925 since explosion -- with the five antennae
equipped with receivers operating in the 3-mm
band (85--105 GHz). The array was in the 750B array configuration on June 30
and July 1, with baselines between 61 and 765 m, and in the H168 array
configuration on August 2, with baselines between 61 and 192 m. Observations
were performed in 2-GHz wide bands centred on frequencies of 93 and 95 GHz.
All three sessions were characterised by stable atmospheric conditions, with
very low precipitable water vapour on June 30 and July 1.

The bandpass calibrator, PKS\,B1921$-$293 was used on June 30 and August 2,
while PKS\,B0637$-$752 was used on July 1. The phase calibrator
PKS\,B0530$-$727 was observed for 1.5 min per 2 min integration time on the
source, while the pointing calibrator PKS\,B0637$-$752 was observed for 2 min
at approximately hourly intervals. Uranus was used as flux density calibrator.
Observations of SNR\,1987A were centred on RA $5^{\rm h}35^{\rm
m}27\rlap{.}^{\rm s}975$, Dec $-69^\circ16^\prime11\rlap{.}^{\prime\prime}08$
(J2000) as from Potter et al.\ (2009).

The data were processed using the {\sc
miriad}\footnote{http://www.atnf.csiro.au/computing/software/miriad} package.
The task {\sc atfix} was first used to correct the system temperatures,
instrumental phases and baseline lengths. The data were examined, and scans
during poor atmospheric phase stability were rejected. To maximise the
signal-to-noise (S/N) ratio, robust weighting was used to invert the
calibrated $uv$ data at both frequencies, with weighting parameter 0.5 (Briggs
1995). A preliminary {\sc clean} model (H\"ogbom 1974) was constructed by
using a small number of iterations (viz.\ 200). The source had sufficient S/N
for phase self-calibration, which was performed with a 2-min solution
interval. Subsequently, deconvolution was performed using the maximum entropy
method (MEM; Gull \& Daniell 1978). For the final image the MEM model of the
combined datasets was restored to a diffraction-limited circular beam with
full-width at half-maximum (FWHM) $0\rlap{.}^{\prime\prime}7$ -- i.e.\ the
synthesized beamsize of the 750B array -- and regridded at a pixel scale of
$0\rlap{.}^{\prime\prime}01$. Figure 1 shows the Stokes-I continuum image; the
off-source r.m.s.\ noise is $\approx0.5$ mJy beam$^{-1}$ and we determine an
integrated flux density of SNR\,1987A of $23.7\pm2.6$ mJy.

To construct a spectral index image, we also reduced observations at 3 cm
performed on 2011 January 25 (Ng et al., in preparation). The reduction
procedures included application of uniform weighting, one iteration of phase
self-calibration over a 5-min solution interval and MEM deconvolution. As for
the 3-mm image, the 3-cm image was restored with a $0\rlap{.}^{\prime\prime}7$
circular beam and regridded at a pixel scale of $0\rlap{.}^{\prime\prime}01$.

\section{Discussion}

%
%
\begin{figure}
\centerline{\psfig{figure=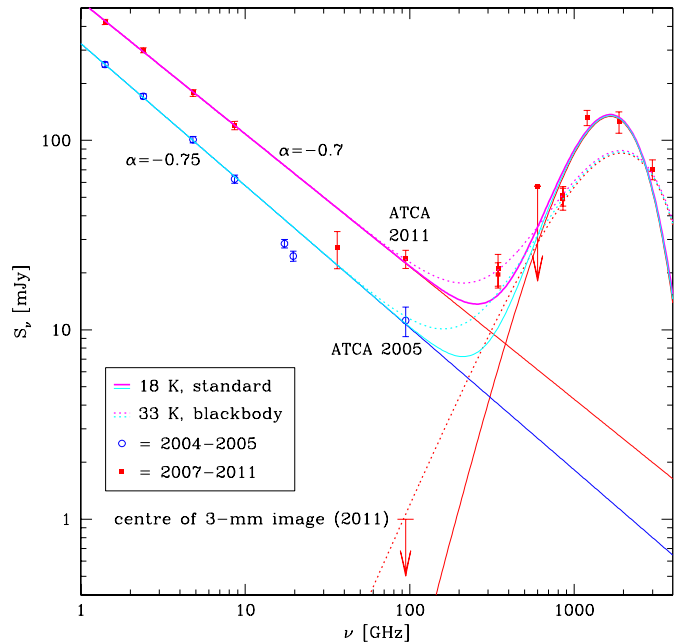,width=87mm}}
\caption[]{IR--radio SED of SNR\,1987A with data from Matsuura et al.\ (2011),
Laki\'cevi\'c et al.\ (2011, 2012), Zanardo et al.\ (2010), and this work.
Power laws are fit to $\lambda>30$ cm, and two different models for dust
emission are plotted (cf.\ Laki\'cevi\'c et al.\ 2012). The upper limit of 1
mJy is plotted for any emission in the centre of the 2011 image at 3 mm in
addition to that extrapolated from the synchrotron power-law.}
\end{figure}

Compared to July 2005 (Laki\'cevi\'c et al.\ 2011), the flux density has
increased $2.1\pm0.4$ times (Fig.\ 2). The new 3-mm measurement lies on the
position expected from extrapolation of a power-law fit to data at
$\lambda>30$ cm of the form $S_\nu\propto\nu^\alpha$, with spectral index
$\alpha=-0.7$, just as the July 2005 measurement at 3 mm (cf.\ Laki\'cevi\'c
et al.\ 2012) obeyed a power law with $\alpha=-0.75$ to the contemporary radio
data. It is odd that the 7- and 12-mm band flux densities are lower than
expected from the power law obeyed by the longer-wavelength data. Could this
result from decay of the synchrotron mechanism? For a decay time
$\tau=17/(B^2\gamma)$ yr (with magnetic field $B$ and relativistic $\gamma$
factor) of $\sim10$ yr, and happening around the characteristic frequency
$\nu=4\times10^6B\gamma^2$ Hz of $\sim10$ GHz, one would derive $B\sim0.1$ G
(which is rather high) and $\gamma\sim150$ (which is typical). However, this
scenario would be hard to reconcile with the ever-increasing synchrotron flux,
and would imply that the 3-mm flux is not due to synchrotron emission.

%
%
\begin{figure}
\centerline{\vbox{
\hspace{7.8mm}\psfig{figure=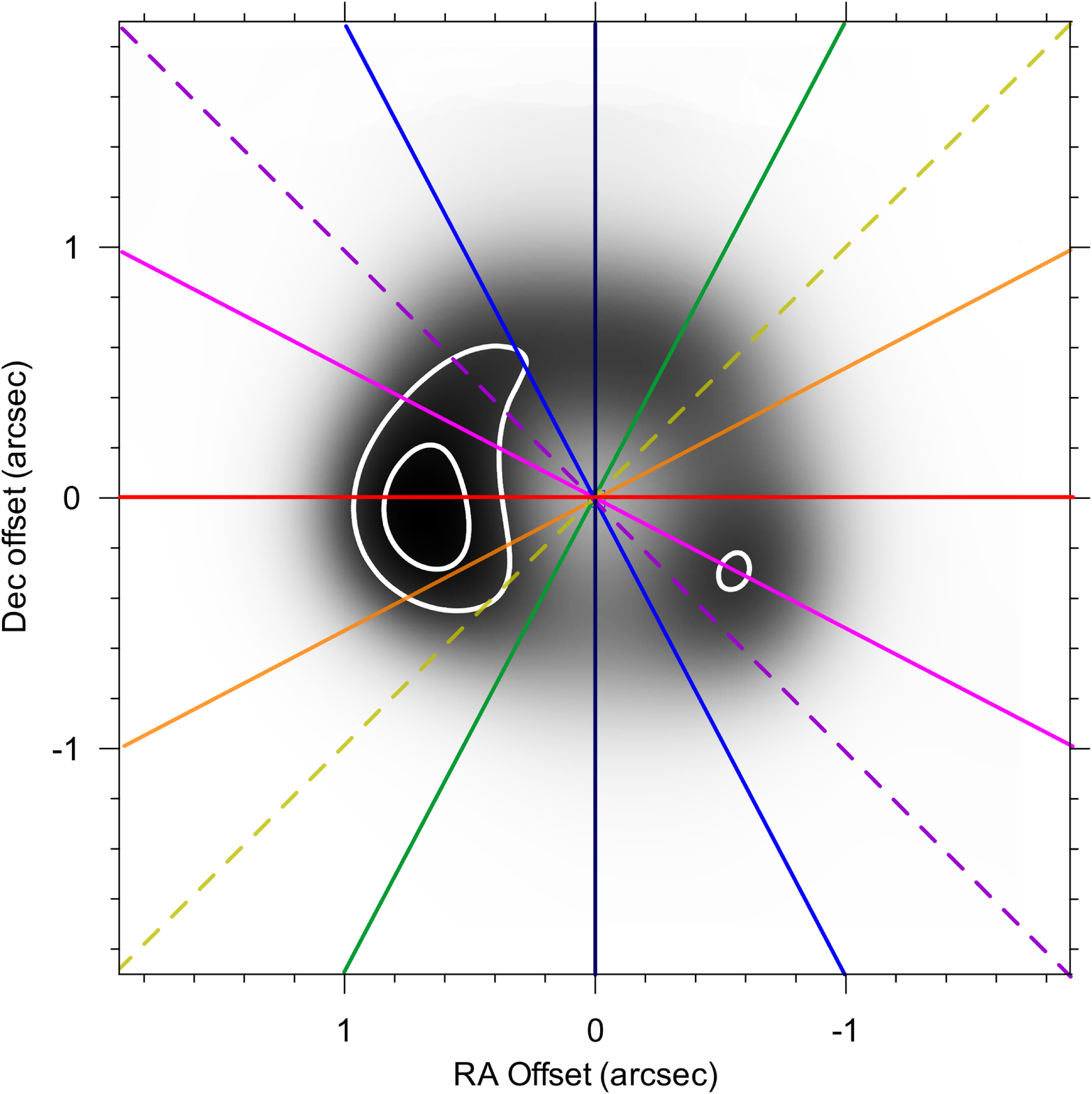,width=67.5mm}
\psfig{figure=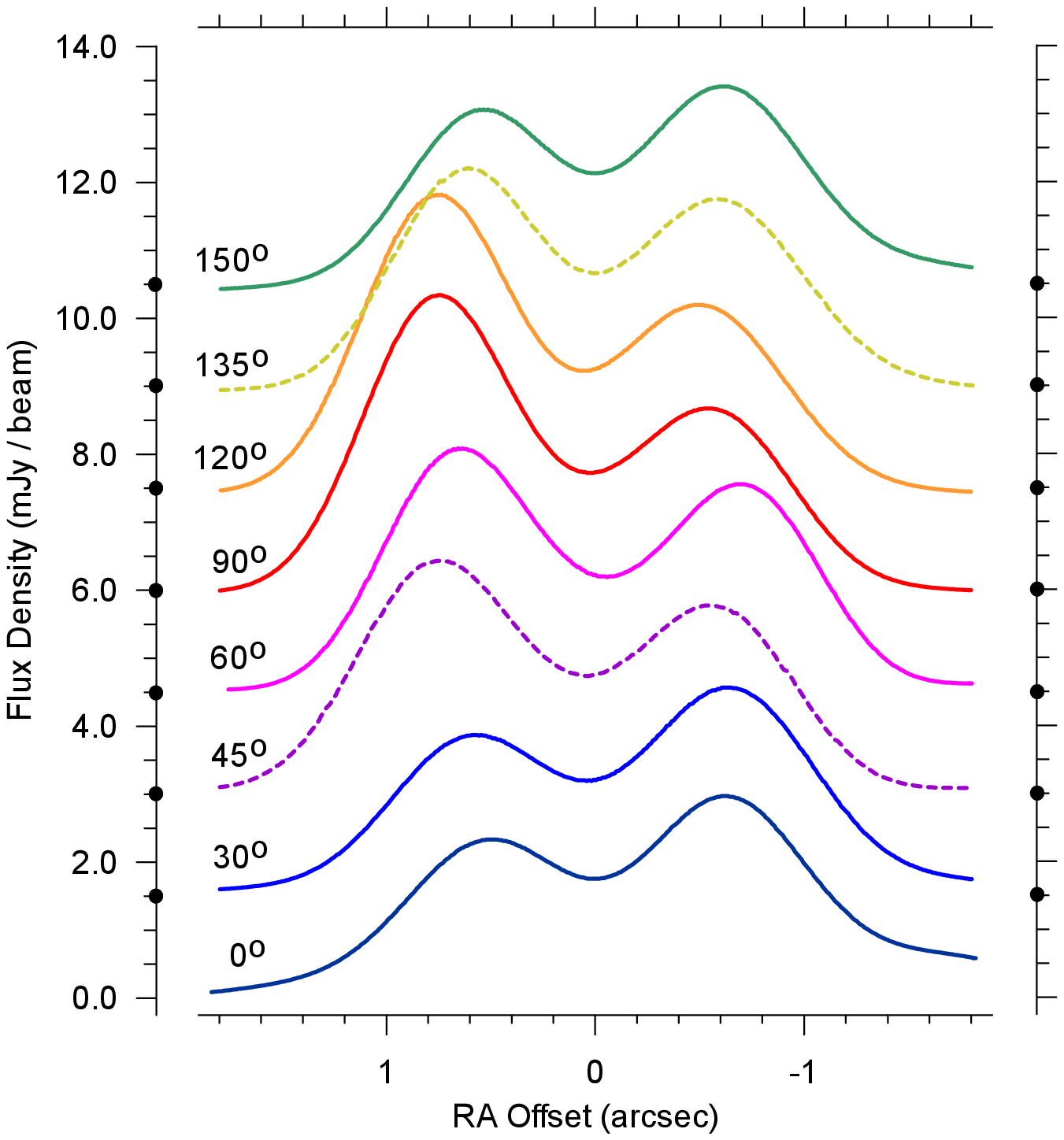,width=79.5mm}
}}
\caption[]{Radial slices through the 3-mm image at 8 position angles. Black
dots on vertical axes indicate the position of the zero for each slice. The
top image is overlain with the 70\% and 90\% flux-density levels (white
contours). The image centre has been shifted to the VLBI position of SN\,1987A
determined by Reynolds et al.\ (1995) [RA $05^{\rm h}35^{\rm m}27\rlap{.}^{\rm
s}968$, Dec $-69^\circ16^\prime11\rlap{.}^{\prime\prime}09$ (J2000)].}
\end{figure}

At $0\rlap{.}^{\prime\prime}7$ resolution the 3-mm emission from SNR\,1987A is
clearly resolved (Fig.\ 1), making it the highest radio frequency at which
this has now been accomplished. The 3-mm image shows similarities with the
images at 7 and 12 mm (Manchester et al.\ 2005; Potter et al.\ 2009) and at 3
cm (Ng et al.\ 2008). The emission is broadly distributed along the equatorial
ring, but unlike the optical images the radio appearance is asymmetric with
the emission peaking in the eastern lobe (Fig.\ 3). The flux ratio between the
eastern and western halves of the image is $\approx1.3$. By fitting an
optically thin shell, this asymmetry was found to be $\approx1.3$ at 7 mm
(Potter et al.\ 2009) and $\approx1.4$ at 3 cm (Ng et al.\ 2008). The maximum
asymmetry with respect to the VLBI position in the 3-mm image occurs near
position angle $90^\circ$ (i.e.\ due east--west), where it reaches a ratio of
$\approx1.6$ (Fig.\ 3).

The flux density in the centre is $\approx1.6$ mJy, but this includes a
contribution from surrounding emission due to beam smearing. The emission
peaks at $\pm0\rlap{.}^{\prime\prime}6$ from the centre, at $\approx3$ mJy
beam$^{-1}$ (Fig.\ 3). The central intensity level resulting from two
identical Gaussians of peak intensity 3 mJy and FWHM
$0\rlap{.}^{\prime\prime}7$ separated by $1\rlap{.}^{\prime\prime}2$ is 0.8
mJy. This leaves room for additional emission in the centre also at a level of
$\approx0.8$ mJy, which may be compared with the 2-$\sigma$ r.m.s.\ noise of
$\approx1$ mJy. Given the crudeness of this estimate we settle on an upper
limit of 1 mJy at 2-$\sigma$ significance. This is below the upper limit of
$\sim2$ mJy placed on contributions from free--free emission to the mm
emission by Laki\'cevi\'c et al.\ (2011). Those authors estimated that the
cold dust ($\sim18$ K) seen at far-IR and sub-mm wavelengths would contribute
only $\sim0.1$ Jy at 3 mm. As the emissivity law is uncertain, Laki\'cevi\'c
et al.\ (2012) considered alternatives, their most extreme model involving
warmer grains ($\sim33$ K) behaving like perfect black bodies at long
wavelengths -- e.g., needles or fluffy aggregates. That model need only
require $<0.01$ M$_\odot$ of dust, and would yield $\sim1$ mJy at 3 mm (Fig.\
2). Thus our current upper limits do not rule out such type of dust (and so
little of it) to be present.

%
%
\begin{figure}
\centerline{\psfig{figure=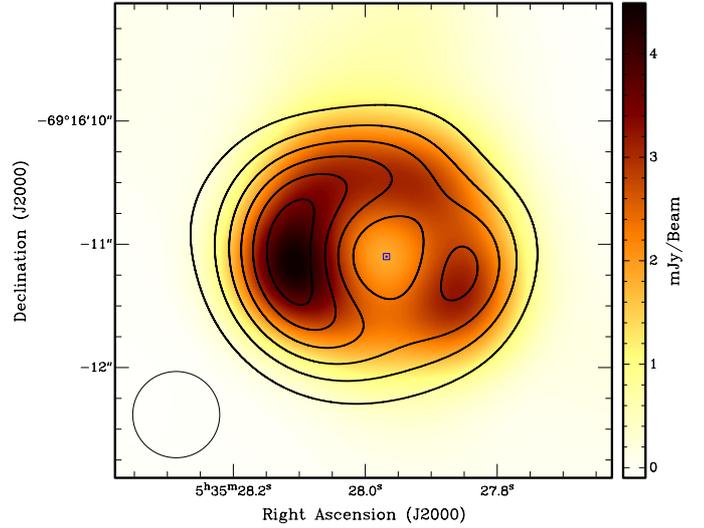,width=90mm}}
\caption[]{The 3-mm image is overlain with the contours of the 3-cm image
derived from observations performed on 2011 January 25 (Ng.\ et al., in
preparation), where the contours correspond to 20\% flux-density intervals.
Both the 3-mm and 3-cm images have been restored with a
$0\rlap{.}^{\prime\prime}7$ circular beam and centred on the VLBI position of
SN\,1987A determined by Reynolds et al.\ (1995) (blue square).}
\end{figure}

%
%
\begin{figure}
\centerline{\psfig{figure=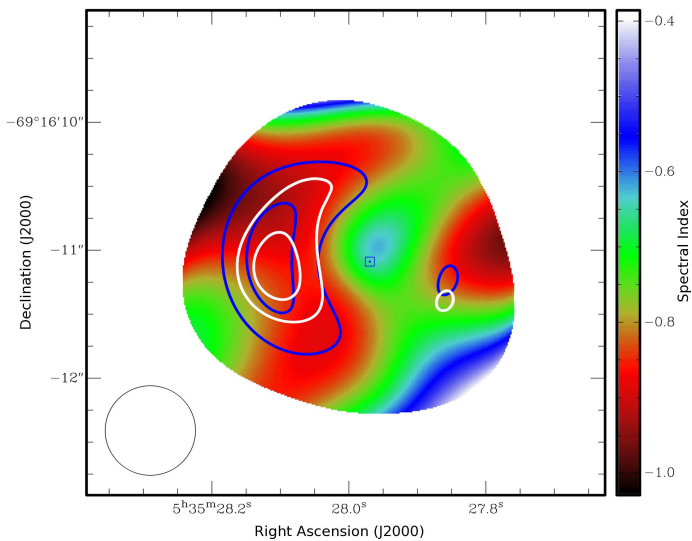,width=90mm}}
\caption[]{3-mm -- 3-cm spectral index image. The spectral index $\alpha$ is
defined as $S_\nu\propto\nu^\alpha$ and was determined from the ratio of the
3-cm image from observations performed on 2011 January 25 (Ng.\ et al., in
preparation) and the 3-mm image from observations around July 2011. Both
images have been restored on a $0\rlap{.}^{\prime\prime}7$ circular beam and
centred on the VLBI position of SN\,1987A determined by Reynolds et al.\
(1995) (blue square). The images are overlain with contours representing the
70\% and 90\% flux density levels (white: 3 mm; blue: 3 cm).}
\end{figure}

%
%
\begin{figure}
\centerline{\psfig{figure=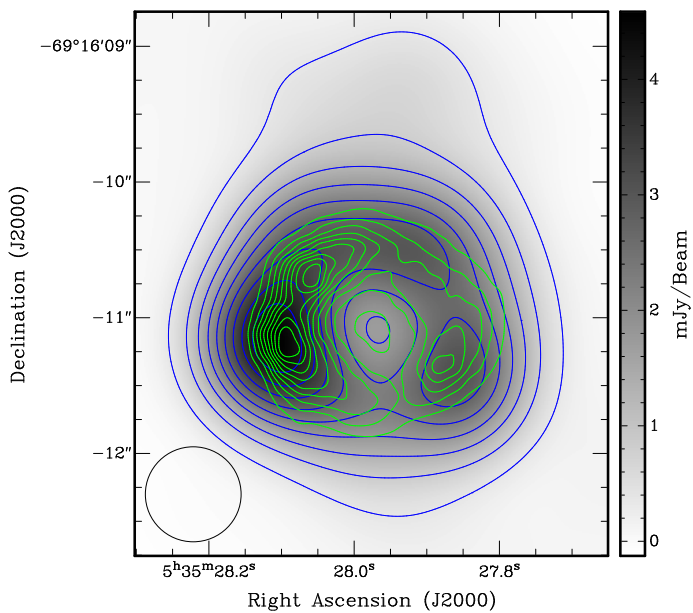,angle=270,width=90mm}}
\caption[]{The 3-mm image of SNR\,1987A from 2011 (greyscale and blue
contours; the beam is plotted in the lower left corner), overlain with (green)
contours of the average of the deconvolved 0.3--8 keV Chandra X-ray images
from October 1999, January 2000 and December 2000 (Burrows et al.\ 2000; Park
et al.\ 2002). Contours are between 10\% and 90\% with 10\% intervals.}
\end{figure}

The 3-mm image is compared with the 3-cm image in Figure 4; these form the
basis for the spectral index image (Fig.\ 5). The spectral index $\alpha$ was
determined from the ratio of the two images where the flux levels exceed
3-$\sigma$ above zero -- still, near the outer edge of the displayed area the
low flux level combined with the inevitable imperfections in the image
reconstruction can easily lead to large variations in the flux ratio between
independent images. We observe the following: (1) the brightest radio emission
is characterised by $-0.95<\alpha<-0.75$, and (2) the centre has a shallower
spectrum, with $\alpha\sim-0.65$, similar to the tentative result obtained at
cm wavelengths (Potter et al.\ 2009).

Diffusive shock acceleration theory (Jones \& Ellison 1991) yields
$\alpha=(1-\gamma)/2$, where $\gamma$ is the power-law index of the momentum
distribution of the accelerated electrons, $N\propto p^{-\gamma}$ and depends
only on the shock compression ratio, $r$, as $\gamma=(r+2)/(1-r)$. Supersonic
shocks are easily established in SNRs, and lead to $1<r\leq4$, i.e.\
$\alpha\leq-0.5$ ($r>4$ is possible for relativistic and/or non-adiabatic
shocks). We conclude that the strongest mm-emission, with
$-0.95<\alpha<-0.75$, arises from moderately-weak shocks with $2.6<r<3.0$. The
central region could signify the emergence of a pulsar wind nebula with strong
shocks ($r\sim3.3$), or some non-synchrotronic emission -- e.g., free-free
emission with $\alpha\sim-0.1$ (indeed, the ejecta are clearly visible in {\it
Hubble} Space Telescope images) or emission from grains with high emissivity
at mm wavelengths (cf.\ Laki\'cevi\'c et al.\ 2012; Wickramasinghe \&
Wickramasinghe 1993), but the spectral index value in the image centre is
uncertain.

There is a striking resemblance between the 3-mm image and the earliest,
deconvolved X-ray images (Fig.\ 6), taken with the Chandra satellite in 1999
and 2000 (Burrows et al.\ 2000; Park et al.\ 2002; Racusin et al.\ 2009), when
the blast wave had just reached the inner rim of the flash-ionized ring. In
particular, the peaks in the mm emission in the east and south--west
correspond to similar peaks in the X-ray emission at that time. The X-ray
images are broad-band, 0.3--8 keV, and were averaged and aligned with the 3-mm
image; the 1999 image was taken in $0^{\rm th}$ order of the High Energy
Transmission Grating, and biased towards higher energies than the 2000 images.
The correspondence between the high-frequency radio structures and X-ray
structures is remarkable given that radio synchrotron emission depends on the
magnetic field strength and is highly anisotropic.

The resemblance between the mm emission and X-rays has diminished since 2000,
as the X-ray images have become dominated by soft X-ray emission from hot gas
behind the forward shock distributed more uniformly along the ring (Racusin et
al.\ 2009; Zhekov et al.\ 2010a). The fact that the synchrotron emission seen
in 2011 is linked to plasma through which the forward shock plowed more than a
decade before suggests that particle acceleration in SNR\,1987A is sustained
at the interface between the ionized plasma of the fast wind from the blue
supergiant progenitor of SN\,1987A, and the denser circumstellar medium from
the preceding red supergiant stage. A magnetised fast wind might also lie at
the origin of the bipolar circumburst morphology -- indeed, the radio emission
extends in the north--south direction (Fig.\ 6; cf.\ the 7-mm images in Potter
et al.\ 2009).

\section{Conclusions}

We have resolved the remnant of SN\,1987A at 3 mm wavelength, at a resolution
of $0\rlap{.}^{\prime\prime}7$. The image is dominated by synchrotron emission
from just inside the equatorial ring, generated by moderately-weak shocks. We
note correspondence between the 3-mm emission and structures in the hard X-ray
emission recorded in 1999 and 2000, i.e.\ more than a decade earlier. These
observations corroborate the idea proposed by Manchester et al.\ (2005) that
the synchrotron emission arises from plasma behind the reverse shock (or
reflected shock; cf.\ Zhekov et al.\ 2010b) which moves relatively slowly
through relatively hot plasma resulting in a moderate Mach number and hence a
moderate compression ratio. We set an upper limit on the emission from any
dust or pulsar wind nebula in the centre, of $\approx1$ mJy (2 $\sigma$).

\begin{acknowledgements}
We thank the staff at Narrabri, in particular our duty astronomer, Tui
Britton, and David Burrows and Eveline Helder who kindly supplied us with the
Chandra images. ML acknowledges studentships from ESO and Keele University.
The Australia Telescope Compact Array is part of the Australia Telescope which
is funded by the Commonwealth of Australia for operation as a National
Facility managed by CSIRO. The Centre for All-sky Astrophysics is an
Australian Research Council Centre of Excellence, funded by grant CE11E0090.
\end{acknowledgements}


\begin{thebibliography}{}
\bibitem[2006]{bouchet} Bouchet P., et al. 2006, ApJ, 650, 212
\bibitem[1995]{briggs} Briggs D.S. 1995, BAAS, 27, 1444
\bibitem[2000]{burrows} Burrows D.N., et al. 2000, ApJ, 543, L149
\bibitem[2010]{dwek} Dwek E., et al. 2010, ApJ, 722, 425
\bibitem[1978]{gull} Gull S.F., Daniell G.J. 1978, Nature, 272, 686
\bibitem[1974]{hogbom} H\"ogbom J.A. 1974, A\&AS, 15, 417
\bibitem[1991]{jones} Jones F.C., Ellison D.C. 1991, Space Sci.\ Rev., 58, 259
\bibitem[2010]{kjaer} Kj{\ae}r K., Leibundgut B., Fransson C., Jerkstrand A.,
Spyromilio J. 2010, A\&A, 517A, 51
\bibitem[2011]{lakicevi1} Laki\'cevi\'c M., van Loon J.Th., Patat F.,
Staveley-Smith L., Zanardo G. 2011, A\&A, 532, L8
\bibitem[2012]{lakicevic2} Laki\'cevi\'c M., van Loon J.Th., Stanke T., De
Breuck C., Patat F. 2012, A\&A Letters, in press
\bibitem[2002]{manchester1} Manchester R.N., Gaensler B.M., Wheaton V.C.,
Staveley-Smith L., Tzioumis A.K., Bizunok N.S., Kesteven M.J., Reynolds J.E.
2002, PASA, 19, 207
\bibitem[2005]{manchester2} Manchester R.N., Gaensler B.M., Staveley-Smith L.,
Kesteven M.J., Tzioumis A.K. 2005, ApJ, 628, L131
\bibitem[2011]{matsuura} Matsuura M., et al. 2011, Science, 333, 1258
\bibitem[2008]{ng} Ng C.-Y., Gaensler B.M., Staveley-Smith L., Manchester
R.N., Kesteven M.J., Ball L., Tzioumis A.K. 2008, ApJ, 684, 481
\bibitem[2002]{park} Park S., Burrows D.N., Garmire G.P., Nousek J.A., McCray
R., Michael E., Zhekov S. 2002, ApJ, 567, 314
\bibitem[2009]{potter} Potter T.M., et al. 2009, ApJ, 705, 261
\bibitem[2009]{racusin} Racusin J.L., Park S., Zhekov S., Burrows D.N.,
Garmire G.P., McCray R. 2009, ApJ, 703, 1752
\bibitem[1995]{reynolds} Reynolds J.E., et al. 1995, A\&A, 304, 116
\bibitem[1993]{wickramasinghe} Wickramasinghe N.C., Wickramasinghe A.N. 1993,
Ap\&SS, 200, 145
\bibitem[2011]{wilson} Wilson W.E., et al. 2011, MNRAS, 416, 832
\bibitem[2010]{zanardo} Zanardo G., et al. 2010, ApJ, 710, 1515
\bibitem[2010a]{zhekov1} Zhekov S.A., McCray R., Dewey D., Canizares C.R.,
Borkowski K.J., Burrows D.N., Park S. 2010a, ApJ, 692, 1190
\bibitem[2010b]{zhekov2} Zhekov S.A., Park S., McCray R., Racusin J.L., Burrows
D.N. 2010b, MNRAS, 407, 1157
\end{thebibliography}
\end{document}